\documentclass[%
reprint,
superscriptaddress,
amssymb,
aps,
prl,
floatfix,
noeprint,
nonote
]{revtex4-2}

\usepackage{graphicx}
\usepackage{physics}
\usepackage{rotating}
\usepackage{amsmath}
\usepackage{bbm}
\usepackage{subfigure}
\usepackage{color}
\usepackage{braket}
\usepackage{tikz}
\usepackage{hyperref}
\hypersetup{colorlinks, 
	linkcolor={blue!75!black!80!yellow},
	citecolor={blue!75!black!80!yellow}, 
	urlcolor={blue!75!black!80!yellow}
	}

\makeatletter \renewcommand\@make@capt@title[2]{%
\@ifx@empty\float@link{\@firstofone}{\expandafter\href\expandafter{\float@link}}%
\sffamily{\textbf{#1}}\@caption@fignum@sep#2 }

\begin{document}

\title{Photon Many-body Dispersion: an Exchange-correlation Functional for Strongly Coupled Light-matter Systems}

\author{Cankut Tasci}
\affiliation{Department of Physics, City College of New York, New York, NY 10031, USA}
\affiliation{Department of Physics, The Graduate Center, City University of New York, New York, NY 10016, USA}
\author{Leonardo A. Cunha}
\affiliation{Center for Computational Quantum Physics, Flatiron Institute, 162 5th Ave., New York, NY 10010,  USA}
\author{Johannes Flick}
  \email[Electronic address:\;]{jflick@ccny.cuny.edu}
\affiliation{Department of Physics, City College of New York, New York, NY 10031, USA}
\affiliation{Department of Physics, The Graduate Center, City University of New York, New York, NY 10016, USA}
\affiliation{Center for Computational Quantum Physics, Flatiron Institute, 162 5th Ave., New York, NY 10010,  USA}

\date{\today}

\begin{abstract}
We introduce an electron-photon exchange-correlation functional for quantum electrodynamical density-functional theory (QEDFT). The approach, photon MBD (pMBD), is inspired by the many-body dispersion (MBD) method for weak intermolecular interactions, which is generalized to include both electronic and photonic (electromagnetic) degrees of freedom on the same footing. We demonstrate that pMBD accurately captures effects that arise in the context of strong light-matter interactions, such as anisotropic electron-photon interactions, beyond single-photon effects, and cavity modulated van der Waals interactions. Moreover, we show that pMBD is computationally efficient and allows simulations of large complex systems coupled to optical cavities.
\end{abstract}

\date{\today}

\maketitle

\textit{Introduction}:
Strong coupling of molecular systems and materials to optical cavities and nanoplasmonic structures has been recently shown to induce modifications in various physical and chemical properties, including ground-state chemical reactivity \cite{thomas2019tilting,ahn2023modification}, excited-state photochemical reactions \cite{hutchison2012m}, among others \cite{bxiang2020i, polak2020m, ojambati2020e, liu2021u}. Despite these promising experimental results, understanding the microscopic origins of these phenomena remains challenging. From a theoretical and computational standpoint, traditional electronic structure methods fall short in capturing electron-photon (e-ph) correlation effects, which become crucial in the strong coupling limit. For instance, e-ph correlation effects are key to elucidate the so-called cavity-induced van der Waals (c-vdW) interactions, which do not decay with the usual $R^{-6}$ dependency on the distance between two atoms, well known for regular van der Waals interactions, but decay with an $R^{-3}$ dependency \cite{haugland2023understanding,philbin2023molecular}. While  wavefunction approaches for electronic structure calculations have been extended to the strong light-matter regime~
\cite{rivera2019,haugland2020coupled,chowdhury2021ring,pavosevic2022cavity,liebenthal2022equation,mallory2022reduced,bauer2023perturbation,cui2024variational,vu2024cavity},~ their steep computational scaling hinders their applicability to more complex systems. 

On the other hand, quantum-electrodynamical density-functional theory (QEDFT) (a generalization of density-functional theory (DFT) to quantum-electrodynamical environments \cite{tokatly2013,ruggenthaler2014}) promises a computationally efficient protocol to simulate these systems. In the QEDFT framework, several approximations~\cite{pellegrini2015,flick2018a,flick2022,schafer2021making,lu2024electronphoton} for the electron-photon exchange-correlation (xc) energy $E_{xc}$ have been proposed, and in particular the single-photon optimized effective potential (OEP)~\cite{pellegrini2015,flick2018a} and a gradient-density approximation (GA)~\cite{flick2022} have been applied to molecular systems. While the OEP approximation exhibits favorable performance for single-photon processes, being an orbital functional entails higher computational costs. The recently proposed gradient-density approximation (GA)~\cite{flick2022} based on the QEDFT fluctuation-dissipation theorem for $E_{xc}$ performs well with reduced computational scaling, enabling simulations of systems with a high number of degrees of freedom. Nevertheless, this approximation falls short in capturing the anisotropic nature of molecules and lacks inclusion of higher-order electron-photon processes, such as multi-photon processes and the cavity-induced van der Waals interaction. Consequently, achieving a deeper understanding of these strongly coupled light-matter systems with many degrees of freedom necessitates the development of improved approximations. In this paper, we introduce a new e-ph xc functional that now allows to accurately and efficiently simulate these complex systems inside an optical cavity while correctly considering anisotropy and higher-order e-ph processes. To overcome previous shortcomings related to anisotropy and higher-order electron-photon interaction terms, we employ the adiabatic-fluctuation dissipation theorem~\cite{flick2022,novokreschenov2023} under the random-phase approximation (RPA) and solve it by connecting to the many-body dispersion (MBD) framework previously employed to describe weak intermolecular (dispersion/van der Waals) interactions~\cite{tkatchenko2012}, particularly the MBD-range separation with self-consistent screening (MBD@rsSCS)~\cite{ambrosetti2014long}.

\textit{Theory}: The general Hamiltonian for a coupled light-matter system in the length-gauge and dipole approximation that describes $N_e$ interacting electrons coupled to $N_p$ photon modes of frequency $\omega_\alpha$ is defined as follows~\cite{tokatly2013, ruggenthaler2014} 
\begin{align}
\label{eq:ham}
    \hat{H}=\hat{T}_\text{e}+\hat{H}_\text{p}+\hat{H}_\text{int}.
\end{align} where
\begin{align}
\label{eq:int}
\hat{H}_\text{int}&=\sum_{i>j}^{N_e}v(\textbf r_{i}, \textbf r_{j}) +\sum_{\alpha=1}^{N_p}\left[-\omega_{\alpha}\hat{q}_{\alpha}\boldsymbol\lambda_{\alpha}\cdot\textbf{R}+\frac{1}{2}(\boldsymbol\lambda_{\alpha}\cdot\textbf{R})^2\right]
\end{align}
Eq. ~\ref{eq:ham} includes the electronic kinetic energy $\hat{T}_{e}$, the photonic Hamiltonian $\hat{H}_{p}=\sum_{\alpha=1}^{N_p}\frac{1}{2}(\hat{p}_{\alpha}^2+\omega^2_{\alpha}\hat{q}_{\alpha}^{2})$ with the photonic momentum and coordinates $\hat{p}_{\alpha}$ and $\hat{q}_{\alpha}$. We collect all interactions in the system in the interaction Hamiltonian $\hat{H}_\text{int}$ (Eq.~\ref{eq:int}),  i.e., 
electron-electron (Coulomb) interaction $v(\textbf r_{i}, \textbf r_{j}) = \frac{e^2}{4\pi\epsilon_{0}}\sum_{i>j}^{N_e}\frac{1}{\abs{\textbf r_{i}- \textbf r_{j}}}
$, and electron-photon interactions via the electric dipole moment $\textbf R=\sum_{i=1}^{N_e}e\textbf{r}_i$ and the photon coordinate $\hat q_\alpha$. Electron-photon interactions also give rise to the so-called dipole self-energy term (DSE) (last term in Eq.~\ref{eq:int}).
The light-matter coupling strength 
$\boldsymbol\lambda_{\alpha}$ is related to the effective cavity volume $V_\text{c}$ by $|\boldsymbol\lambda_{\alpha}| =1/\sqrt{\epsilon_0 V_\text{c}}$~\cite{climent2019plasmonic}.

The adiabatic-connection fluctuation-dissipation theorem can be used to obtain xc energies ($E_\text{xc}$) within the QEDFT framework~\cite{flick2022,novokreschenov2023}

\begin{align}
\label{eq:ad-con}
E_\text{xc}&= -
\frac{1}{2\pi} \int^1_0 d\gamma  \int_0^\infty d\omega \, \text{Tr} \left[\mathcal G(i\omega) \chi_\gamma(i\omega) \right],
\end{align}
where $\text{Tr}[..]$ implies spatial integration over $\textbf r$ and $\textbf r'$~\footnote{We are omitting spatial dependencies for clarity.}. Eq.~\ref{eq:ad-con} only contains two ingredients, (1) the electronic response function $\chi_\gamma$ that depends on the dimensionless parameter $\gamma$ to interpolate between the non-interacting system at $\gamma=0$ and the physical Hamiltonian at $\gamma=1$ and (2) the effective electronic propagator $\mathcal G(i\omega)$.

The interacting response function $\chi_{\gamma=1}=\chi$ can be connected to the Kohn-Sham response function $\chi_0=\chi_{\gamma=0}$ via a Dyson equation in the frequency domain that reads~\cite{flick2019} \footnote{We drop dependency on $\omega$ for convenience.}
\begin{align}
\chi = \chi_0 + \chi_0 \left[f_\text{Hxc} + f_\text{pxc}\right]\chi
\label{eq:chi}
\end{align}
where $f_\text{Hxc}$ is the Hartree-exchange-correlation kernel originating from the Coulomb interaction~\cite{ullrich2012time} and $f_\text{pxc}$ is the kernel due to electron-photon interaction~\cite{flick2019}. 

We further find for the effective electronic propagator 
\begin{align*}
\mathcal G(i\omega) = v(\textbf{r},\textbf{r}')+\sum_{\alpha=1}^{N_p} \frac{\omega^2}{\omega_\alpha^2 -\omega^2}\left(\boldsymbol\lambda_\alpha\cdot\textbf r \right)\left( \boldsymbol\lambda_\alpha\cdot\textbf{r}'\right) 
\end{align*}

that contains the Coulomb interaction $v$ between the electrons and an effective electronic dipole-dipole interaction that is mediated by the cavity modes.

We emphasize in passing that Eq.~\ref{eq:ad-con} describes the total xc energy of the interacting light-matter system, which for $|\boldsymbol\lambda_\alpha| \rightarrow 0$ reduces to the regular DFT electronic xc energy~\cite{hesselmann2011}. 

We can now divide the xc energy of Eq.~\ref{eq:ad-con} into two parts, the exchange energy $E_\text x$ and the correlation energy $E_\text c$ that are defined as follows
\begin{align}
\label{eq:ex}
E_\text{x}&= -
\frac{1}{2\pi}  \int_0^\infty d\omega \, \text{Tr} \left[\mathcal G(i\omega) \chi_0(i\omega) \right]\\
\label{eq:ec}
E_\text{c}&= -
\frac{1}{2\pi} \int^1_0 d\gamma  \int_0^\infty d\omega \, \text{Tr} \left[\mathcal G(i\omega) \left(\chi_\gamma(i\omega)-\chi_0(i\omega)\right) \right]
\end{align}

We emphasize here that earlier approximate QEDFT functionals, the OEP approach of Ref.~\cite{pellegrini2015, flick2018a} and the GA formulation ~\cite{flick2022} have been based on approximating the electron-photon part in the exchange energy $E_x$ in Eq.~\ref{eq:ex} that describes one-photon processes, while the electronic part in $E_x$ corresponds to the exact-exchange energy expression of regular DFT~\cite{hesselmann2011}.

In the following, we include higher-order correlation effects explicitly in the QEDFT functional. We will apply the random-phase approximation (RPA)~\cite{hesselmann2011,novokreschenov2023} that neglects exchange-correlation components in $f_\text{Hxc}$ and $f_\text{pxc}$. Thus, in Eq.~\ref{eq:chi} we set $f_\text{Hxc} = f_\text{H}$ and $f_\text{pxc} = f_\text{p}$. Doing so, yields 
\begin{align*}
\chi = \chi_0 + \chi_0 \mathcal G \chi.
\end{align*}

Using RPA, we can now perform the $\gamma$ integration in Eq.~\ref{eq:ad-con} explicitly to obtain the correlation energy $E_c$ alternatively as a sum~\cite{tkatchenko2013}
\begin{align*}
E_\text{c,RPA}&= -
\frac{1}{2\pi}  \int_0^\infty d\omega \,  \sum_{n=2}^\infty\frac{1}{n}\text{Tr}\left[\left(\mathcal G(i\omega) \chi_0(i\omega)\right)^n\right]
\end{align*}
It is insightful to look at the lowest order ($n=2$), which reads
\begin{widetext}
\begin{align}
\label{eq:2nd}
E_\text{c,RPA} &= -
\frac{1}{4\pi}  \int_0^\infty d\omega \int \int \int \int d\textbf{r} d\textbf r' d\textbf r'' d\textbf r'''\sum_{ijkl}\alpha_{ij}(\textbf r, \textbf r', i\omega)  \alpha_{kl}(\textbf r'', \textbf r''', i\omega) \nonumber\\
&\times \left[ T_{jk}(\textbf r', \textbf r'')T_{li}(\textbf r''', \textbf r) + \sum_\alpha\frac{2\omega^2}{\omega^2+\omega_\alpha^2}T_{jk}(\textbf r', \textbf r'')  \lambda^{(\alpha)}_l \lambda^{(\alpha)}_i + \sum_{\alpha,\beta}\frac{\omega^2}{\omega^2+\omega_\alpha^2}\frac{\omega^2}{\omega^2+\omega_\beta^2}\lambda^{(\alpha)}_j \lambda^{(\alpha)}_k  \lambda^{(\alpha)}_l \lambda^{(\alpha)}_i\right] + ...
\end{align}
\end{widetext}

In Eq.~\ref{eq:2nd}, we have introduced the electronic polarizability $\alpha_{ij}$ via $\chi(\textbf r, \textbf r', i\omega) = -\nabla_{r_i} \nabla_{r'_j}\alpha_{ij}(\textbf r, \textbf r', i\omega)$ and the dipole interaction tensor ${T}_{ij}(\textbf r, \textbf r') = \nabla_{r_i} \nabla_{r'_j}v(\textbf r, \textbf r')$.
For the lowest order in Eq.~\ref{eq:2nd}, we identify three terms. The first term $\sim T^2$ is the well-known van der Waals (vdW) interaction and behaves as $\sim R^{-6}$ for the distance $R$ of two (neutral) atoms~\cite{wu2002,grimme2010,hermann2017}. The second term $\sim \lambda^2 T$ and the last term $\sim \lambda^4$
have been identified in Ref.~\cite{haugland2023understanding,philbin2023molecular} as cavity-induced van der Waals (c-vdW) interaction that leads to a $\sim \lambda^2 R^{-3}$ behavior and  a collective cavity-induced energy that is distance-independent, respectively.

\textit{Method:} To formulate a QEDFT density functional that is able to capture higher-order electron-photon correlation effects, such as c-vdW interactions, we now combine ideas developed to treat dispersion interactions using DFT (in particular the many-body dispersion (MBD) framework~\cite{tkatchenko2012,tkatchenko2013,bucko2013,ambrosetti2014long}), linear-response QEDFT~\cite{flick2018a} and the QEDFT fluctuation-dissipation theorem~\cite{flick2022,novokreschenov2023}. 

First, we will apply the MBD framework to describe the long-range electronic correlations. We assume that we can divide the electronic correlation part, i.e. the contribution due to $v(\textbf{r},\textbf{r}')$, of Eq.~\ref{eq:ec} into a short-range and a long-range part.~\footnote{Here we assume that the short-range part can be efficiently treated by standard DFT xc functionals.}. Next, we assume that the long-range part of $E_c$ electronic system can be effectively described by different atomic fragments for each of the $N_{a}$ atoms that we obtain via the Hirshfeld partitioning scheme~\cite{tkatchenko2009,tkatchenko2012}. An effective dipole polarizability is then associated with each atom as
$\alpha_{i}(i\omega)=\frac{\alpha_{i}}{1+\omega^{2}/\omega_{i}^{2}}$, where $\alpha_{i}$ is the static ($\omega=0$) polarizability of atom $i$, and $\omega_i$ is a characteristic excitation frequency. To obtain $\alpha_i$ the self-consistent screening (SCS) equation is solved~\cite{Oxtoby1975,tkatchenko2012,ambrosetti2014long} to include short-range (sr) range-separated self-consistent screening (rsSCS). As a last step the MBD Hamiltonian that describes a system of coupled quantum harmonic oscillators, one for each atom, is solved to obtain the electronic correlation energy $E_c$. We refer the reader to Refs.~\cite{ambrosetti2014long,hermann2017} and the Supplemental Material (SM) for more details on the MBD approach.

In the following we will now extend the MBD framework to solve Eq.~\ref{eq:ad-con} for correlated electron-photon systems and denote the new approach as photon MBD (pMBD). Specifically, we extend the MBD Hamiltonian by including additional dimensions corresponding to the individual photon modes. A detailed derivation can be found in the SM. This extension is reminiscent of earlier generalizations of electronic structure methods, ie. the QEDFT Casida equation~\cite{flick2018a} and the light-matter force-constant matrix~\cite{bonini2022}.
Thus, we have to solve the following Hamiltonian that consists of $3 N_a+N_p$ coupled quantum harmonic oscillators and describes the $N_a$ atoms and $N_p$ photon modes
\begin{align}
\label{eq:rpa-ham}
       \hat{H}_\text{pMBD}&=\frac{1}{2}\sum_{i=1}^{N_a}\sum_{a=1}^{3}\left(-\nabla^{2}_{ia}+\omega^2_{ia}\chi_{ia}^{2}\right)\nonumber\\
       &+\frac{1}{2}\sum_{i,j=1}^{N_a}\sum_{a,b=1}^3\omega_{ia}\omega_{jb}\sqrt{\alpha_{ia}\alpha_{jb}}\chi_{ia}T_{\text{LR},ij}^{ab}\chi_{jb}\nonumber\\
       &+ \frac{1}{2}\sum_{\alpha=1}^{N_p}\hat{p}_{\alpha}^{2} +\frac{1}{2}\sum_{\alpha=1}^{N_{p}}(\omega_{\alpha}\hat{q}_{\alpha}-\sum_{i=1}^{N_a}\sum_{a=1}^{3}\lambda_{\alpha a}\omega_{ia}\sqrt{\alpha_{ia}}\chi_{ia})^{2}.
\end{align}
Here, the $N_a$ atoms are described by mass-weighted displacements from equilibrium $\chi_{i}=\sqrt{m_i}\xi_{i}$ and parameterized by their effective frequencies $\omega_{ia}$, charges $e_{i}$ and polarizabilities $\alpha_{ia} = e_{i}^2/(m_{i}\omega_{ia}^2)$. In Eq.~\ref{eq:rpa-ham}, $T_{\text{LR},ij}^{ab}$ describes the long-range part of the dipole-dipole interaction tensor~\cite{ambrosetti2014long}. The light-matter interaction is then described by coupling of the atomic dipole moments $\mu_{i}=e_{i}\xi_{i}$ and the photon displacement coordinates $\hat{q}_\alpha$. While solving the SCS equation for $\alpha_{ia}$~\cite{Oxtoby1975,tkatchenko2012,ambrosetti2014long} leads to anisotropic polarizabilies, we use isotropic polarizabilies $\alpha_i = \sum_{a=1}^3 \alpha_{ia}/3$ to recover regular MBD energies in the limit $|\boldsymbol \lambda_\alpha |= 0$. We emphasize that anisotropic effects are included in Eq.~\ref{eq:rpa-ham} via $T_{\text{LR},ij}^{ab}$.

Under these assumptions, $E_\text{xc}$ can then be expressed as the difference between the interacting and non-interacting energies. Thus, we can define the pMBD exchange-correlation energy 
\begin{align}
\label{eq:rpa-xc}
    E_\text{xc,pMBD}=\frac{1}{2}\sum_{k=1}^{3N}\Omega_{k}-\frac{1}{2}\sum_{i=1}^{N}\sum_{a=1}^{3}\omega_{ia}-\frac{1}{2}\sum_{\alpha=1}^{N_{p}}\omega_{\alpha}    
\end{align}
where $\Omega_{k}^{2}$ are the eigenvalues of the Hamiltonian defined in Eq.~\ref{eq:rpa-ham}, $\omega_{ia}$ the effective atomic frequencies and $\omega_{\alpha}$ the cavity frequencies. We note that Eq.~\ref{eq:rpa-xc} can now be combined with regular DFT functionals (non-selfconsistently) that describe the electronic exchange and (short-range) correlation contributions. To get the total xc energy, both contributions will have to be combined~\footnote{In practice, the parameter $\beta$ that enters the Fermi-type damping function that separates $T_{ia,jb}$ into a long-range and short-range part has to be chosen depending on the electronic DFT functional. We use here the standard MBD values, i.e. $\beta = 0.83$ for PBE and $\beta=0.85$ for PBE0, see Ref.~\cite{ambrosetti2014long}.}.

\begin{figure}[t]
\centering
\includegraphics[width=1.0\linewidth]{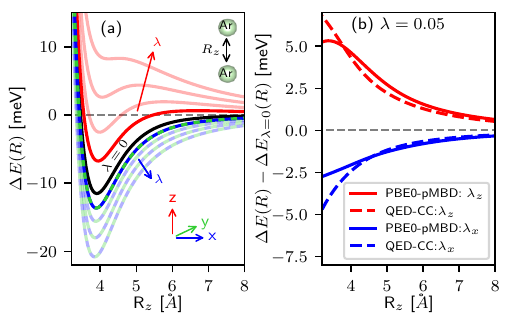}
\caption{Interaction energy $\Delta E$ comparison of Ar dimer outside (black) and inside the cavity along different polarization directions. In (a), we show the interaction energy $\Delta E$ for different distances between two argon atoms, and polarization along $z$ (red), $x$ (blue) and $y$ (green) direction. In (b), we compare cavity energy results between PBE0-pMBD and QED-CC. For both methods, $\Delta E (R_z) = E (R_z) - E (R_z = 25 \AA)$}
\label{fig:01}
\end{figure}

\textit{Application:} In the remainder of this letter, we will exemplify the pMBD  approach with three different examples of coupled molecule-cavity systems: a prototypical van der Waals system, the Ar dimer; a benzene dimer complex, and a bilayer H-terminated graphene flake consisting of 144 carbon and hydrogen atoms. In all three examples, we couple the electronic system to a single cavity mode and employ the PBE0 functional~\cite{perdew1996rationale} as the electronic xc functional, given the success of the MBD method with the PBE0 functional~\cite{tkatchenko2012}. We will refer to this approach as PBE0-pMBD. For the case of the Ar dimer, we will compare the pMBD approach to accurate polaritonic coupled-cluster (QED-CC) calculations~\cite{haugland2020coupled,pavosevic2022cavity}. We refer to the SM for the numerical details on the pMBD and QED-CC computations. Unless specifically noted, all results were obtained by coupling the molecular systems to a single cavity mode with frequency $\omega_\alpha = 2$eV and coupling strength magnitude $|\boldsymbol{ \lambda}_\alpha|=0.05$ a.u.

We first present the results for an Ar dimer strongly coupled to a single cavity mode (Fig.~\ref{fig:01}). We place two argon atoms parallel to the $z$-axis and increase the distance $R_z$ as shown in the inset of Fig.~\ref{fig:01} (a). In Fig. \ref{fig:01} (a), we show the interaction energy $\Delta E(R_z)$ as a function of the separation of the two Ar atoms distance. The black solid line depicts the density functional theory (DFT) results outside the cavity, i.e. PBE0-MBD or $\lambda=0$. Results using the PBE0-pMBD with electron-photon coupling strength of $|\boldsymbol \lambda_\alpha|=0.05$ a.u. are shown with the solid red, solid blue, and solid green lines, which represent cavity polarizations (inset) in $z$, $x$, and $y$ direction respectively. Results for cavity polarization in $x$ and $y$ are identical. Additionally we show higher coupling strength, up to $|\boldsymbol \lambda_\alpha|=0.1$ a.u. in their respective shaded color. We find that the PBE0-pMBD approach is capable of capturing the anisotropic nature of the electron-photon interaction, as well as correctly describe the c-vdW interactions~\cite{haugland2023understanding}. We refer to the SM for an analysis of the different $R^{-6}$ and $R^{-3}$ contributions. We further find a change in sign for the c-vdW corrections. Notably, in the z-direction (red), the c-vdW energy contribution is repulsive, while in the $y$ (green) and $x$ (blue) directions, it is attractive, leading to stronger ($x$/$y$) or weaker ($z$) overall interactions between the two Ar atoms. If the electron-photon coupling becomes very strong ($|\boldsymbol \lambda_\alpha|=0.1$), the total interaction becomes overall repulsive and no stable minima is present. Fig. \ref{fig:01} (b) presents a comparison between the pMBD approach and QED-CC. Here, we indicate the difference in $\Delta E$ inside and outside ($|\boldsymbol \lambda_\alpha| =0$) the cavity. Overall, we find excellent agreement between the pMBD method and the QED-CC, validating the accuracy of the pMBD approach.

\begin{figure}[t]
\centering
\includegraphics[width=1.0\linewidth]{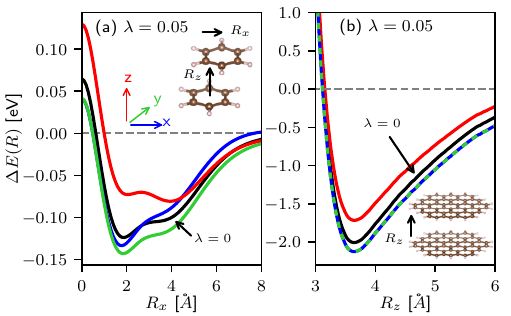}
\caption{Interaction energy $\Delta E$ comparison of benzene dimer complex and graphene flake outside (black) and inside the cavity with different polarization directions $z$ (red), $x$ (blue) and $y$ (green) direction. In (a), we fix $R_{z}=3.3 \AA$ and change $R_{x}$ values. In (b), we keep two graphene flakes parallel to each other along the $xy$ plane, and change the value of $R_{z}$.}
\label{fig:02}
\end{figure}

In the second example (Fig. \ref{fig:02} (a)), we study parallel-displaced benzene dimer (PD-benzene) structure in an optical cavity. 
The PD-benzene structure has two displacement variables $R_{x}$ and $R_{z}$, representing horizontal and vertical displacements, respectively as shown in the inset in Fig.~\ref{fig:02} (a). In this case, we fix $R_{z}=3.3 \AA$, and vary $R_{x}$. In Fig. \ref{fig:02} (a), the black solid line represents simulation outside the cavity, i.e., $|\boldsymbol\lambda_\alpha|=0$, which corresponds to a PBE0-MBD simulation. Here, the minimum of the potential well is located around $R_{x}=2\AA$. Placing the system in an optical cavity, with $|\boldsymbol\lambda_\alpha|=0.05$, we are able to capture the sign change of the c-vdW contribution to the interaction energy $\Delta E$ in different polarization directions. Interestingly, in the polarization along the $z$-direction (red), the cavity-induced effects change the position of the minimum in the potential energy surface quite significantly, from $R_{x}=2\AA$ to $R_{x}=4\AA$. Cavity polarization along $x$ (blue) and $y$ (green) direction leads to stronger attractive interactions between the benzene molecules.

Finally, owing to its lower computational cost, we apply the pMBD approach to investigate a larger system, a graphene flake dimer coupled to a single cavity mode. 
Each flake consists of 54 carbon atoms terminated by 18 hydrogen atoms, i.e. in total 144 atoms. Due to its size, this system is too large to be simulated with QED-CC methods. In Fig. \ref{fig:02} (b), we show the results of the PBE0-pMBD simulations for the interaction energy $\Delta E$. The inset shows the setup: the two graphene flakes are placed parallel to each other in the $x$-$y$ plane, and we change the distance $R_{z}$ in $z$-direction in each simulation. Due to the symmetry considerations, we find similar interaction energies if the cavity is polarized along $x$ (blue) or $y$ (green) directions. In these cases, the interaction energy inside the cavity becomes more attractive compared to the $|\boldsymbol\lambda_\alpha|=0$ (black) case. In contrast if the cavity polarization is oriented along $z$-direction (out-of-plane), the interaction energy becomes less attractive than the $|\boldsymbol\lambda_\alpha|=0$ case. We also note that the changes in the interaction energy are in the order of several 100s meV.

\textit{Summary and conclusion}: In this letter, we introduce the photon MBD (pMBD)  QEDFT functional. The pMBD functional is the first xc functional within the QEDFT framework that incorporates higher-order electron-photon processes (many-photon processes and c-vdw effects), captures the anisotropic nature of complex molecular structures, and allows the computationally efficient simulation of complex systems coupled to cavity modes. We demonstrate and validate the accuracy of the pMBD functional by comparing it with the expensive, but accurate QED-CC method for the Ar dimer system. Furthermore, we find significant changes on the potential-energy minima for a benzene dimer inside an optical cavity, depending on the direction of polarization. Additionally, we present results for stacked graphene flakes inside an optical cavity, which, due to its size, is out-of-reach for QED-CC methods.
This work now opens a path for an accurate and efficient ab-initio description of complex light-matter systems at the interface of quantum optics and chemistry.

\textit{Acknowledgements}: We acknowledge funding from NSF via grant number EES-2112550 (NSF Phase II CREST Center IDEALS) and startup funding from the City College of New York. All calculations were performed using the computational facilities of the Flatiron Institute. The Flatiron Institute is a division of the Simons Foundation.

\bibliography{refs} 

\end{document}


\title{Supplemental Material for: \\ ``Photon Many-body Dispersion: an Exchange-correlation Functional for Strongly Coupled Light-matter Systems"}

\author{Cankut Tasci}
\affiliation{Department of Physics, City College of New York, New York, NY 10031, USA}
\affiliation{Department of Physics, The Graduate Center, City University of New York, New York, NY 10016, USA}
\author{Leonardo A. Cunha}
\affiliation{Center for Computational Quantum Physics, Flatiron Institute, 162 5th Ave., New York, NY 10010,  USA}
\author{Johannes Flick}
  \email[Electronic address:\;]{jflick@ccny.cuny.edu}
\affiliation{Department of Physics, City College of New York, New York, NY 10031, USA}
\affiliation{Department of Physics, The Graduate Center, City University of New York, New York, NY 10016, USA}
\affiliation{Center for Computational Quantum Physics, Flatiron Institute, 162 5th Ave., New York, NY 10010,  USA}

\date{\today}

\date{\today}

\maketitle

\subsection{Numerical details of DFT calculations}
In all calculations, we employ the PySCF python package \cite{sun-pyscf2020} for all density-functional theory (DFT) calculations. We employ the PBE0 \cite{Perdew1996PBE, perdew1996rationale} functional and the aug-cc-pVDZ basis set~\cite{kendall1992electron}. For most of the calculations, we use the default real-space grid settings defined in PySCF. The only exception is the Ar-dimer, where we use the highest available real-space grid settings (grid level 9), due to spatial fluctuations in the long-range distance limit. To find the Hirshfeld volume ratios, we utilize the VDW python package~\footnote{Accessible at \url{https://github.com/ajz34/vdw}}. Finally, for MBD@rsSCS calculations, we employ the libmbd library~\cite{Hermann2023}. All the pMBD calculations are performed in a non self-consistently. To construct the benzene dimer, we use the experimental gas phase geometry of a single benzene available from NIST~\cite{johnson2015nist}. In the case of the graphene dimer, we relax a single graphene flake using PySCF and the PBE0 functional until energy differences become  smaller than 10$^{-6}$  Hartree.

\subsection{Numerical details of coupled-cluster calculations}

QED coupled cluster theory (QED-CC) is a generalization of the popular CC family of methods to the polaritonic regime \cite{haugland2020coupled,pavosevic2021polaritonic}.
In QED-CC, the cluster operator that enters the exponential ansatz includes  purely electronic ($T_{x0}$), purely photonic ($T_{0y}$) and mixed electron-photon ($T_{xy}$) excitations.
Energies and amplitudes are obtained as in regular CC theory, \emph{i.e.} through projections into the appropriate subspaces \cite{haugland2020coupled}.
As introduced in Ref.~\citenum{pavosevic2021polaritonic}, we only considered a single photonic mode, and truncated such operators to include up to double excitations in both electronic, photonic and mixed sectors, as showed in Eq.~\ref{eq:qedcc}. 
This model is known as QED-CCSD-22.
\begin{equation}\label{eq:qedcc}
\begin{split}
        T &= T_{10} + T_{20} + T_{01} + T_{02} + T_{11} + T_{12} + T_{21} + T_{22} \\
        T_{0y} &= \left( b^\dagger\right)^y \\
        T_{1y} &= \sum_{ia}\left( b^\dagger\right)^y a^\dagger i \\
        T_{2y} &= \frac{1}{4}\sum_{ijab}\left( b^\dagger\right)^y a^\dagger b^\dagger j i
\end{split}
\end{equation}

QED-CC calculations were performed using a modified version of the code available in Ref.~\citenum{pavosevic2022cavity}. In this work, we introduce the frozen core approximation to QED-CC, in which electronic excitations out of chemically inactive core orbitals (such as the 1s levels of carbon) are not included in the cluster expansion. QED-CC amplitudes and energies were converged up to 10$^{-8}$ a.u. using conventional DIIS algorithm to solve the CC amplitude equations \cite{scuseria1986accelerating}. Energies outside of the cavity were obtained at the regular CCSD level with the same numerical thresholds. All CCSD and QED-CCSD-22 calculations were carried out with the aug-cc-pVDZ basis set~\cite{kendall1992electron}.

\subsection{Derivation of photon MBD (pMBD) Hamiltonian}

To derive the pMBD Hamiltonian described in the main text, we expand the MBD Hamiltonian~\cite{tkatchenko2012,ambrosetti2014long} to include the photonic subspace. 

We first follow the steps described in Ref.~\cite{donchev2006} for the coupled fluctuated dipole model (CFDM), which is the basis of the MBD Hamiltonian~\cite{tkatchenko2012,ambrosetti2014long} and then add the photonic Hamiltonian and the electron-photon interaction terms. In the CFDM model, atoms are treated as three dimensional quantum harmonic oscillators, each with atomic mass $m_{i}$ and effective excitation frequency $w_{i}$. All atoms are coupled by a dipole-dipole interaction potential $T_{ij}^{ab}$. The CFDM Hamiltonian for a system of $N_a$ atoms can then be written as~\cite{donchev2006}
\begin{equation}
    \begin{split}
    {H}_\text{CFDM}=&-\frac{1}{2}\sum_{i=1}^{N_a}\sum_{a=1}^{3}\frac{\nabla^{2}_{ia}}{m_{i}}+\frac{1}{2}\sum_{i=1}^{N_a}\sum_{a=1}^{3}m_{i}\omega_{ia}^{2}\xi^{2}_{ia}\\
    &+\sum_{i>j=1}^{N_a}\sum_{a,b=1}^{3}e_{i}e_{j}\xi_{ia}T_{ij}^{ab}\xi_{jb}
\end{split}
\end{equation}
where $\xi_{ia}$ describes the displacement of the $a$th ($a=1,2,3$) component of the $i$th atom from its equilibrium position.

Inside an optical cavity, we can now define the photon CFDM (pCFDM) Hamiltonian as
\begin{equation}
    \begin{split}
H_\text{pCFDM}&=H_\text{CFDM}+\frac{1}{2}\sum_{\alpha=1}^{N_{p}}p_{\alpha}^{2}\\&+\frac{1}{2}\sum_{\alpha=1}^{N_{p}}(\omega_{\alpha}q_{\alpha}-\sum_{i=1}^{N}\sum_{a=1}^{3}\lambda_{\alpha a}e_{i}\xi_{ia})^{2}.
 \end{split}
\end{equation}

Here, $p_{\alpha}$ and $q_{\alpha}$ are photon momentum and coordinate, respectively, $\omega_{\alpha}$ is cavity mode frequency, and $\lambda_{\alpha a}$ the electron-photon coupling strength, all as defined in the main text. Matter and photons are now coupling via the instantaneous dipole moment $\mu_{ia} = e_{i}\xi_{ia}$. After introducing the mass-weighted coordinates $\chi_{ia}=\sqrt{m_{i}}\xi_{ia}$, and using the atom resolved electric polarizability $\alpha_{ia}=e_{i}^2/(m_{i}\omega_{ia}^2)$, we find the pMBD Hamiltonian as presented in the main text.

\subsection{Treatment of dipole-dipole interaction tensor $T^{ab}_{ij}$ and polarizabilities $\alpha_{ia}$}

To calculate the range-separation of the dipole-dipole interaction tensor $T^{ab}_{ij}$ necessary to avoid double counting, we use the regular MBD procedure described in \cite{ambrosetti2014long}, which we summarize in the following. The Coulomb interaction potential due to two charged atom-centered quantum harmonic oscillators (QHO) located at $r_{i}$ and $r_{j}$, and separated by $r_{ij}=\left| r_{i}-r_{j} \right|$ is given by
\begin{equation}
v(r_{ij})=\frac{{\text{erf}}(r_{ij}/\sigma_{ij})}{r_{ij}}
\end{equation}
where $\sigma_{ij}=\sqrt{\sigma_i^2+\sigma_j^2}$, and $\sigma_{i}=\left( \sqrt{2/\pi}\alpha_{i}/3 \right)^{1/3}$ is the Gaussian width of the $i$th QHO. Therefore, the dipole-dipole interaction tensor can be defined as
\begin{equation}
\label{eq:dtensor}
T^{ab}_{ij}=\partial_{r_i^{a}}\partial_{r_j^{b}}v(r_{ij})
\end{equation}
where $r_{i}^{a}$ represents the $a$th Cartesian coordinate of $r_i$. The short-range behavior of Eq. \ref{eq:dtensor} is not compatible for calculation of long-range correlation energy, since short-range effects are already included via the SCS equation for the polarizability~\cite{ambrosetti2014long}. Thus, the tensor is separated into two parts, short-range and long-range. The short-range part is defined as
\begin{equation}
T_{\text{SR},ij}^{ab}=(1-f(r_{ij}))T_{ij}^{ab}
\end{equation}
where $f(r_{ij})$ is a Fermi-type damping function, defined as
\begin{equation}
f(r_{ij})=\frac{1}{1+\exp{(-a(r_{ij}/S_{\text{vdW}}-1))}}
\end{equation}
where $a=6$, and $S_{\text{vdW}}=\beta(R^{i}_{\text{vdW}}-R^{j}_{\text{vdW}})$, $R^{i}_{\text{vdW}}$ is the van der Waals radii of the $i$th atom. Thus, the long-range part can be written as
\begin{align}
T^{ab}_{\text{LR},ij}&=f(r_{ij})T^{ab}_{ij}\\
&=f(r_{ij})\frac{-3r_{ij}^ar_{ij}^b+r_{ij}^2\delta_{ab}}{r_{ij}^5}
\end{align}
$T_{\text{LR},ij}^{ab}$ enters then the pMBD Hamiltonian as defined in the main text. {The short-range part enters the SCS equation~\cite{tkatchenko2009} via
\begin{align}
\alpha^{\text{SCS}}_{ia} = \alpha^{0}_{ia} - \alpha^{0}_{ia}\sum_{jb}T_{\text{SR},ij}^{ab}\alpha^{\text{SCS}}_{jb},
\end{align}
where $\alpha^{0}_{ia} = \delta_{lm}\alpha_i^\text{TS}$ are the TS polarizabilities~\cite{tkatchenko2009}
and $\alpha^{\text{SCS}}_{ia}$ are the self-consistently screened polarizabilies. As stated in the main text, in the pMBD Hamiltonian, the isotropic average $\alpha_{ia} = \sum_{b}\alpha^\text{SCS}_{ib}/3$ is employed.}
For more details, we refer the reader to Refs.~\cite{ambrosetti2014long,hermann2017}.

\subsection{Cavity modified van der Waals interactions (c-vdW)}

As discussed in the main manuscript, the optical cavity can modulate intermolecular interactions and change their asymptotic behavior. In the case of dispersion or van der Waals interactions, the emergence of a term that depends on the inverse cubic power of the distance ($R^{-3}$) between two molecules has been reported both numerically and analytically \cite{haugland2023understanding,philbin2023molecular}.

This cavity-modulated $R^{-3}$ dependence is captured with accurate wavefunction methods, such as polaritonic extensions of coupled cluster theory (QED-CC). In this section, our goal is to demonstrate that the photon MBD (pMBD) approach developed in the QEDFT framework is capable of accurately modelling such cavity-modulated interactions.

\begin{figure}
    \centering
    \includegraphics[width=1.0\linewidth]{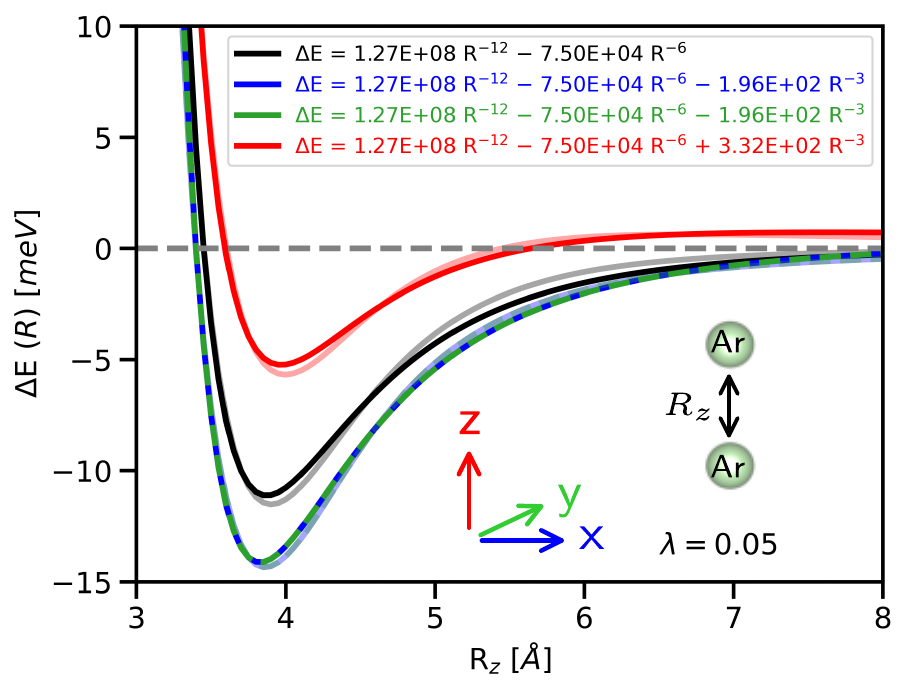}
    \caption{Cavity-modulated c-vdW interactions in the Ar dimer ($\lambda = 0.05$ a.u.). The transparent lines indicate the original data, whereas the solid ones represent fits to the potential energy surfaces, highlighting the emergence of a $R^{-3}$ dependent term that can stabilize or destabilize the interaction between the Ar atoms depending on the polarization of the cavity.}
    \label{fig:si_r3}
\end{figure}

Fig.~\ref{fig:si_r3} illustrates the emergence of a $R^{-3}$ dependent term for the interaction energy inside the cavity for the Ar dimer for $|\boldsymbol\lambda_\alpha|=0.05$ a.u. Such a term is crucial to explain the differences in both asymptotic behavior and modification of the minimum of the potential energy curve. In this particular example, we find that, for cavities polarized in the $x$ and $y$ direction (blue and green curves in Fig.~\ref{fig:si_r3}, the $R^{-3}$ helps stabilize the interaction between the two Ar atoms with respect to the results outside the cavity (black curve in Fig.~\ref{fig:si_r3}). However, such a term destabilizes the interaction for a cavity polarized in the $z$ direction, leading to a shallower well and a repulsive tail at longer ranges of the potential energy curve (red curve in Fig.~\ref{fig:si_r3}).

Finally, we can assess if our pMBD approach recovers the behavior predicted by perturbation theory \cite{haugland2023understanding} for the ration between the prefactors of the c-vdW interaction with respect to different cavity polarizations. According to perturbation theory, the c-vdW interaction is given for a strictly isotropic system by

\begin{equation}
    E_\textrm{c-vdW} \left( R \right)= \frac{\lambda^2 C_6}{R^3} \left( cos^2\left( \theta \right) -\frac{1}{3}\right)
\end{equation}

where $\lambda$, $C_6$, $R$ and $\theta$ are, respectively, the cavity coupling strength, the van der Waals $C_6$ coefficient, the distance between the two monomers, and the angle between the cavity polarization and the displacement vector between the monomers. In our setup, this displacement vector is in the $z$ direction. Hence, for a cavity polarized in the $z$ direction, we have $\theta = 0$, whereas for a cavity polarized in the $x$ (or $y$) direction, $\theta = \frac{\pi}{2}$. This results in a ratio of 2 between the c-vdW contributions for a cavity polarized in the z direction compared to one polarized in the x/y direction. We can estimate this from our QEDFT results by computing

\begin{equation}
    r = \abs{\frac{\Delta E_{\lambda_z} - \Delta E_{\lambda=0}}{\Delta E_{\lambda_x} - \Delta E_{\lambda=0}}}
\end{equation}

At $R_z = 4.0\AA$, we can estimate this ratio to be 2.18, which is close to the value predicted by the perturbative analysis and indicates that our pMBD approach is able to accurately capture c-vdW interactions. Differences can be attributed to the inclusion of anisotropic effects in the pMBD framework.

\bibliography{refs} 